\documentclass[prl,aps,floatfix,superscriptaddress,footinbib,twocolumn,10pt,English]{revtex4}

\usepackage{amssymb,amsmath}
\usepackage{graphicx}
\usepackage[percent]{overpic}
\usepackage{overpic}
\usepackage{color}
\usepackage[colorlinks=true]{hyperref}
\usepackage{enumitem}
\usepackage{bm}

\newcommand{\sS}{{\bf s}}
\newcommand{\hS}{{\bf h}}

\newcommand{\nS}{{\bf n}}
\newcommand{\omegaS}{\Omega}
\newcommand{\gammaS}{\Gamma}
\newcommand{\phiS}{\Phi}
\newcommand{\AS}{{\bf A}}

\begin{document}  

\title{Black hole spectroscopy with coherent mode stacking} 
{
\author{Huan Yang}
\affiliation{Department of Physics, Princeton University, Princeton, New Jersey 08544, USA.}
\author{Kent Yagi}
\affiliation{Department of Physics, Princeton University, Princeton, New Jersey 08544, USA.}
\author{Jonathan Blackman}
\affiliation{TAPIR, Walter Burke Institute for Theoretical Physics,
California Institute of Technology, Pasadena, California 91125, USA.}
\author{Luis Lehner}
\affiliation{Perimeter Institute for Theoretical Physics, Waterloo, Ontario N2L 2Y5, Canada}
\affiliation{CIFAR, Cosmology \& Gravity Program, Toronto, ON M5G 1Z8, Canada}
\author{Vasileios Paschalidis}
\affiliation{Department of Physics, Princeton University, Princeton, New Jersey 08544, USA.}
\author{Frans Pretorius}
\affiliation{Department of Physics, Princeton University, Princeton, New Jersey 08544, USA.}
\affiliation{CIFAR, Cosmology \& Gravity Program, Toronto, ON M5G 1Z8, Canada}
\author{Nicol\'as Yunes}
\affiliation{eXtreme Gravity Institute, Department of Physics, Montana State University, Bozeman, Montana 59717, USA}
\date{\today}

\begin{abstract} 
The measurement of multiple {\em ringdown modes} in gravitational waves from binary black hole mergers will allow for testing fundamental properties of black holes in General Relativity, and to constrain modified theories of gravity. 
To enhance the ability of Advanced LIGO/Virgo to perform such tasks, we propose a {\em coherent mode stacking method} to 
search for a chosen target mode within a collection
of multiple merger events. We first rescale each signal so that the target mode in each of them has 
the same frequency, and then sum the waveforms constructively.
A crucial element to realize this coherent superposition is to make use of \emph{a priori} information
extracted from the inspiral-merger phase of each event.
To illustrate the method, we perform a study with simulated events targeting the
$\ell=m=3$ ringdown mode of the remnant black holes. We show that this
method can significantly boost the signal-to-noise ratio of the
collective target mode compared to that of the single loudest
event. Using current estimates of merger rates we show
that it is likely that advanced-era detectors can measure this
collective ringdown mode with one year of coincident data gathered at
design sensitivity.

\end{abstract}

\maketitle 

{\noindent}{\bf Introduction}.~The recent detection of gravitational
waves (GWs) emitted during the coalescence of binary black
holes~\cite{Abbott:2016blz,Abbott:2016nmj} marked the beginning of the
era of gravitational wave astronomy, a feat that heralds a boom of
scientific discoveries to come. GWs not only provide a
new window to our universe, they also offer a unique opportunity to
test General Relativity (GR) in the dynamical and highly non-linear
gravitational regime~\cite{Yunes:2013dva,TheLIGOScientific:2016src,Yunes:2016jcc,TheLIGOScientific:2016pea,lasky2016detecting}. One
celebrated prediction of GR is the uniqueness, or ``no-hair'' property
of vacuum black holes (BHs) ~\cite{israel,1971PhRvL..26..331C,hawking-uniqueness,1975PhRvL..34..905R,cardoso2016testing}:
{\em all} isolated BHs are described by the Kerr family of
solutions, each uniquely characterized by only its mass and
spin~\footnote{An astrophysical environment is not a pure vacuum,
  though it is {\em expected} that any ambient matter/radiation/charge
  about an aLIGO merger event will have an insignificant effect on the
  spacetime dynamics and corresponding GW
  emission. Also see discussions in \cite{PhysRevD.89.104059}.}. This property has many wide-ranging consequences, the
two most relevant here being (a) that the spacetime of an isolated
binary black hole (BBH) inspiral is uniquely characterized by a small,
finite set of parameters identifying the two BHs in the
binary and the properties of the orbit, and (b) that this same set of
parameters uniquely determines the merger remnant and the full
spectrum of its quasinormal mode (QNM) ringdown waveform.

This latter point forms the basis of \emph{black hole spectroscopy}, where measurements of multiple ringdown modes are used to test this no-hair property. The idea is as follows. If the no-hair property holds, a measurement of the (complex) frequency of one QNM can be inverted to find a discrete set of possibilities for the spherical harmonic $(\ell,m)$ plus overtone number $n$ of the mode, and the BH mass $M$ and spin parameter $a = |\vec{S}|/M^2$, where $\vec{S}$ is the BH spin angular momentum.  
However, if we have \emph{a priori} information about the objects that merged to form the perturbed BH, then we also have information about the dominant $(\ell,m,n)$ QNM, and the measurement of its complex frequency then provides information about the mass and spin of the perturbed object. The measurement of any additional QNM frequencies then overconstrains this mass and spin measurement, providing independent tests of the no-hair property. 
Naturally, the results of such tests can then be leveraged to place constraints on (or to detect) non-Kerr BHs in modified gravity theories, exotic compact objects, the presence of 
exotic/unexpected matter fields, etc. (e.g.~\cite{Dreyer:2003bv,Berti:2005ys,Kamaretsos:2011um,Gossan:2011ha,Jackiw,CSreview,yunespretorius,kent-CSBH,1992PhLB..285..199C,Kanti:1995vq,yunesstein,Pani:2009wy,Yunes:2011we,pani-quadratic,Ayzenberg:2014aka,Brito:2013xaa}).

In fact, aLIGO has already given us a ``zeroth-order'' test of
the no-hair property from event GW150914:~the {\em inspiral only} portion of
the signal was matched to a best-fit numerical relativity template,
giving an estimate of the mass and spin of the remnant,
{\em and} informing that the waveform shortly after peak
amplitude should be dominated by the fundamental harmonic of the
$(\ell,m)=(2,2)$ QNM (``22-mode'' for short); this was consistent
with the independently measured properties of the {\em post-merger} signal~\cite{Abbott:2016nmj}. 
More stringent tests of the no-hair property of the final BH 
require observation of sub-leading QNMs~\footnote{See e.g.~\cite{Nakano:2015uja} for a consistency test of GR with the dominant ringdown mode only.}. This is challenging using
{\em individual} merger events given how weak 
these sub-leading modes are relative to the primary mode~\cite{Berti:2016lat,Bhagwat:2016ntk}.
For example, GW150914 has a ringdown signal-to-noise ratio (SNR) of $\approx 7$,
but a ringdown SNR upwards of $45$ would 
have been needed to detect the first sub-leading QNM~\cite{Gossan:2011ha,Berti:2007zu}. Thus detection
of such modes {\em in individual events} will require third generation GW detectors, as
even a loud GW150914-like event at aLIGO's design sensitivity would have a ringdown SNR of $\approx 20$~\cite{Berti:2016lat}.
On the other hand, many such events are expected after years of operation,
leading us to consider how the information from {\em multiple} 
detections could be used to extract faint signals from a population of events.

Here then, we propose a way to {\em coherently} combine (or ``stack'')
multiple, high total SNR (low ringdown SNR) binary BH coalescence events, to boost the
detectability of a {\em chosen} secondary QNM mode. An earlier study
in~\cite{Meidam:2014jpa} considered a similar problem, though their
approach effectively amounted to an incoherent assembly of ringdown
signals, where, all else being equal, one expects $N^{1/4}$ scaling of
the SNR for $N$ events, compared to $N^{1/2}$ for a coherent method
(see Supplemental Material for more details). Key to achieving
coherent stacking is using information gleaned from the inspiral
portion of each event to predict the relative phases and amplitudes of
the ringdown modes excited in the remnant.

\vspace{2mm}

{\noindent}{\bf Signal stacking}.~Given a set of BBH coalescence
observations, we first select the loudest subset, here taken to
consist of the signals with ringdown SNR in the primary 22-mode alone
of $\rho_{22} > 8$.  Based on the studies
in~\cite{Berti:2005ys,Berti:2007zu,London:2014cma,Bhagwat:2016ntk,PhysRevD.75.124018,PhysRevD.76.064034} the
33-mode is typically one of the next loudest ringdown modes.
Therefore, we concentrate on the 33-mode as a target for our analysis,
although the methodology presented here is generally applicable to
other modes, as well as other features common to a population of GW
events.  Similar to the analysis in
\cite{Berti:2016lat,Bhagwat:2016ntk}, we use the two-mode
approximation to describe each detected {\em ringdown} signal $s_j(t)$
:
\begin{align}\label{eq:decom}
s_j = n_j + h_{22,j}+ h_{33,j}\,,
\end{align}
where the subscript $j$ refers to the $j$th event,  $n_j$ is the corresponding detector noise,
and $h_{\ell m,j}$ is a ringdown mode of the form (for $t>0$)
\begin{align}
\label{eq:h-of-t}
h_{\ell m,j}(t) = A_{\ell m,j} e^{-\gamma_{\ell m,j} t} \sin (\omega_{\ell m,j}\, t - \phi_{\ell m,j})\,.
\end{align}
For each ringdown mode, $(\omega_{\ell m,j} +i \gamma_{\ell m,j})$ is its complex
frequency, $A_{\ell m,j}$ its real amplitude, and $\phi_{\ell m,j}$ its constant phase offset.

Next, each \emph{entire} $j$th signal is fitted to inspiral-merger-ringdown (IMR) waveform models
in GR to accurately extract certain binary parameters that characterize the inspiral (e.g.~the individual
masses and spins)\footnote{Parameter uncertainties scale inversely with SNR, and thus they will likely be
smaller than uncertainties in parameter extraction with GW150914. Such uncertainties should
  have a small effect on the final BH mass and spin measurements, as was
  the case for GW150914~\cite{Abbott:2016blz}}.
Using this, we can compute the QNM frequencies, phase offsets and amplitudes for all modes as
expected in GR (the extrinsic parameters, such as the polarization and
inclination angles do not affect the phase difference between the 22-
and $\ell \ell$-modes ($\ell >2$)~\footnote{If one wishes to search for a subdominant ringdown mode
with $\ell \neq m$, one needs to take into account uncertainties of 
the extrinsic parameters in the phase difference between
the 22 mode and the $\ell m$ mode. For example, based on extrinsic parameter 
uncertainties for stellar-mass BH binaries
in Fig.~11 of~\cite{Veitch:2009hd} with a network of three interferometers
(and neglecting correlations among parameters), 
we found that such uncertainties introduce an error
on the phase difference of the 22 and 21 mode as $\sim 0.2 \times (20/\rho)$ rads, 
which is smaller than the error from intrinsic parameter uncertainties that we used
in our analysis.}, 
as we discuss in the Supplemental
Materials). This is a key ingredient of our coherent mode stacking, as
we need to properly align the phase offsets $\phi_{33,j}$ {\em and}
frequencies $\omega_{33,j}$ of the targeted modes to achieve optimal
improvement in SNR relative to a single event analysis.

To perform the alignment, out of the set of $N$ events, we {\em
  arbitrarily} pick one (e.g.~the $i$th one) as the base case, and
shift/rescale all others to give the same expected secondary mode
phase offset $\phi_{33,i}\equiv \phi_{33}$ and frequency
$\omega_{33,i}\equiv\omega_{33}$.
Specifically, we scale and shift each signal in time via $\sS_j(t)
\equiv s_j(t/\alpha_j + \Delta_j)$, with
$\alpha_j\equiv\omega_{33,j}/\omega_{33}$ and $\Delta_j \equiv
(\phi_{3 3,j}-\phi_{33})/\omega_{33,j}$.

We are now ready to combine the individual signals. For convenience we
work in the frequency domain, denoting the Fourier transform of a function $g(t)$ by $\tilde{g}(f)$. The Fourier transform of Eq.~\eqref{eq:h-of-t} is given by~\cite{Berti:2007zu}
\begin{align}
\tilde h_{\ell m,j} (f) = A_{\ell m,j} \frac{\omega_{\ell m,j} \cos \phi_{\ell m,j} -(\gamma_{\ell m,j} -i\, \omega) \sin \phi_{\ell m,j}}{\omega^2_{\ell m,j}-\omega^2+\gamma_{\ell m,j}^2-2 i\, \omega\, \gamma_{\ell m,j}}
\label{eq:model}
\end{align}
with $\omega=2\pi f$ the angular Fourier frequency. In the frequency domain, the secondary mode alignment of Eq.~(\ref{eq:decom}) is achieved via $\tilde{\sS}_j(f)\equiv \alpha_j e^{i\omega\Delta_j\alpha_j} \tilde{s}_j(\alpha_j f)$.
We then sum up these phase- and frequency-aligned signals to obtain our composite signal: $\tilde {\sS} = \sum_j c_j \tilde {\sS}_j \equiv \tilde \nS+\tilde \hS_{22}+\tilde \hS_{33}$, where the identification of $\tilde \nS$, $\tilde \hS_{22}$ and $\tilde \hS_{33}$ is obvious, and we describe later how to optimize the choice of weight constants $c_j$. If the frequencies and phase offsets are known exactly, $\tilde \hS_{33}$ contains a single oscillation frequency $\omega_{33}$, and $\tilde \hS_{22}$ contains a family of modes with (rescaled) frequencies 
$\in (0.623,2/3)\omega_{33}$ as the dimensionless BH spin $a$ ranges from $0-1$~\cite{Berti:2005ys, PhysRevD.86.104006}.

\vspace{0.2cm}

{\noindent}{\bf Parameter uncertainty}.~Equation~\eqref{eq:decom}
decomposes a measured event into a true underlying signal and detector
noise. The rescaling we have just described makes crucial use of
parameters of the signal during the IMR phase, which can only be
estimated to within some uncertainty, and this will introduce what we call
``parameter estimation noise'' $n_h$, that we will add to the
composite signal $\tilde \sS$. We investigate the role of this
uncertainty here, leaving detailed derivations of some of the
conclusions to the Supplemental Materials. 

Parameter uncertainty produces two main sources of parameter
estimation noise $n_{h}$. The first arises from subtracting an
imperfectly-estimated $\hS_{22}$ from the data. This noise source 
has frequency components quite close to the scaled 
frequencies $\omega_{22,j}$, which (in relative terms, when comparing to $\omega_{33,j}$)
are far from $\omega_{33}$; the latter is the frequency at which $\tilde{\bf{h}}_{33}$ peaks, and thus,
the impact of this noise source on $\rho_{33}$ is small. The second noise
source is due to the imperfect scaling and alignment of the $33$ mode, 
which is resonant at frequency $\omega_{33}$. 

Let us denote any variable with a prime as the maximum likelihood estimator, i.e. $Y'=Y+\delta Y$, with $Y$ the true (scaled or not) value, and $\delta Y$ the corresponding uncertainty in its estimation.
With this, the time domain, estimated composite GW signal is
\begin{align}\label{eqcom2}
&\hS_{22}' = {\rm Im} \left \{\sum_j  \AS'_{22,j}\,  e^{i (\Lambda'_{22,j} t-\phiS'_{22,j})} \right \}\,, \nonumber \\
 & \hS_{33}'=  {\rm Im} \left \{ e^{i (\omega_{33} t - \phi_{33})}\sum_j  \AS'_{33,j}\, 
 e^{-\gammaS'_{33,j}t+i (\delta \omegaS_{33,j} t-\delta \phiS_{33,j})} \right \},
\end{align}
where $\omegaS_{\ell m,j} + i \gammaS_{\ell m,j} \equiv (\omega_{\ell
  m,j} + i \gamma_{\ell m,j})/\alpha_j \equiv \Lambda_{\ell m,j}$ and
$\phiS_{\ell m,j}\equiv \phi_{\ell m,j}-\Delta_j\, \omega_{\ell m,j}$
are the scaled frequencies and phase offsets respectively, and we have
absorbed the $c_j$ coefficients into rescaled amplitudes $\AS_{\ell
  m,j}$.  
The parameter estimation noise for each $(\ell,m)$ mode is $n_{h_{\ell m}} = \hS'_{\ell m}-\hS_{\ell m}$, 
which is approximately given by
\begin{align}\label{eq:paran}
& n_{h_{\ell m}} \approx {\rm Im} \left \{ \sum_j [   \delta \AS_{\ell m,j}\,  e^{i (\Lambda_{\ell m,j} t-\phiS_{\ell m,j})}  \right . \nonumber \\
 &+ \left . \AS_{\ell m,j}\, e^{i (\Lambda_{\ell m,j} t-\phiS_{\ell m,j})} 
 ( e^{i (\delta \Lambda_{\ell m,j} t-\delta \phiS_{\ell m,j})}-1) ] \vphantom{\sqrt{\frac12}}  \right \}\,.
 \end{align}
In the subsequent analysis we {\em assume} that $\delta \AS, \delta \Lambda$
and $\delta \phiS$ are independent, normal random variables in the probability
space of $\nS$~\footnote{We neglect correlations between these variables. 
We have verified that after normalizing the Fisher matrix of these variables
calculated from propagation of errors of inspiral parameters such that
all diagonal components are unity, off diagonal components
are smaller than 44\% for a GW150914-like event. We leave it to
future work to investigate this more thoroughly.}. 

We are unaware of any \emph{closed-form, analytic} formula in the
literature that describes parameter uncertainties given the SNR of a
particular detection, even when the waveform model is known
analytically. Let us then assume one characterizes the data with an
inspiral-merger-ringdown model, where the ringdown contains the 22-
and 33-modes. These ringdown modes depend (of course) on the ringdown
parameters and the underlying gravitational theory governing the
dynamics, though in our analysis we are assuming GR as the theory and
hence they fundamentally depend on the parameters of the inspiral. The
uncertainty in the inspiral parameters depends inversely on the total
SNR $\rho$ of the observation, as can be shown via a simple Fisher
analysis, which then also provides the uncertainty of the ringdown
parameters to within a factor given by the propagation of errors from
the inspiral to ringdown parameters. Guided by an estimate of this
propagation factor as outlined in the Supplemental Material, together
with aLIGO's parameter estimation errors for event
GW150914~\cite{Abbott:2016blz,TheLIGOScientific:2016pea}, we estimate
the variance of mode parameter uncertainties as
$\sigma_{\phiS_{ii,j}} = 0.3 \times (20/ \rho_j)$ rads ($i=1,2$) and
use the QNM
frequency formula and the formula for $A_{ii,j}$ to propagate the mass
uncertainty of event GW150914 to obtain estimates for $\sigma_{\Lambda_{ii,j}}$ 
and $\sigma_{A_{ii,j}}$. 

\vspace{0.2cm}

{\noindent}{\bf Hypothesis Testing}.~With the combined signals, we perform a Bayesian hypothesis test~\cite{Berti:2007zu} to derive the conditions 
of detectability of the 33-mode. In particular, we want to test the following two nested hypotheses:
\begin{align}
&\mathcal{H}1: \quad \tilde y \equiv \tilde \sS -\tilde \hS_{22}= \tilde \nS+A\, \tilde \hS_{33}\,, \nonumber\\
&\mathcal{H}2: \quad \tilde y \equiv \tilde \sS -\tilde \hS_{22} = \tilde \nS.\label{eq:hyp}
\end{align}
For convenience we have introduced an overall amplitude factor $A$ such that when $A \neq 0$ the $33$-mode is non-zero, and vice-versa. The probability that the observed data is consistent with ${\cal{H}}_{1}$ is 
 \begin{align}
 P_{A} \propto {\rm exp} \left [ -\int^\infty_0 df \frac{2 |\tilde y-A \tilde \hS_{33}|^2}{S_n}\right ]\,,
 \label{eq:prop-PA}
 \end{align}
with $S_n = \sum_j c_j^2 S_{n_j} (\alpha_j f)  \alpha_j$ the one-sided and shifted noise spectrum (with $S_{n_j}$ the unscaled
detector noise spectral density for each detection).

With the above probability function, we can derive the maximum likelihood estimator for $A$ and then perform a Generalized Likelihood Ratio Test (GLRT)~\cite{Berti:2007zu}. As we explain in detail in the Supplemental Material, parameter uncertainties shift the mean and expand the variance of the distribution of the likelihood ratio between the two hypotheses. The former effectively reduces the 33 mode to 
\begin{align}
\label{eq:h33bar}
{\bf H}_{33} &= \left[1 + \frac{1}{2} \left( \left \langle \frac{\langle n_{h_{33}} | n_{h_{33}} \rangle}{\langle {\hS}_{33} | {\hS}_{33} \rangle} \right \rangle  -  \left \langle \frac{\langle {\hS}_{33} | n_{h_{33}} \rangle^2}{\langle {\hS}_{33} | {\hS}_{33} \rangle^2} \right \rangle \right) \right] \langle \hS_{33} \rangle\,,
\end{align}
(see the Supplemental Materials for the definition of the inner product $\langle | \rangle$ and explicit form of $\langle \hS_{33} \rangle$), while the latter directly reduces the SNR of the 33 mode by $\sqrt{1 + \sigma_p^2}$ where $\sigma^2_{\rm p}$ is the variance of ${\langle {\hS}_{33} |  n_{h_{22}} -  n_{h_{33}} \rangle }{ \left[ \langle {\hS}_{33}| {\hS}_{33} \rangle\right]^{-1/2}}$ .
Thus, the requirement to favor ${\cal{H}}_{1}$ over ${\cal{H}}_{2}$ is 
\begin{align}\label{eq:rho33}
\rho_{33} \equiv \frac{ \sqrt{\langle {\bf H}_{33} |{\bf H}_{33}\rangle}}{\sqrt{1+\sigma^2_{\rm p}}} \ge \rho_{\rm crit}\,,
\end{align}
where $\rho_{\rm crit}$ is related to the false-alarm rate $P_{\rm f}$ and detection rate $P_{\rm d}$ in the GLRT. If we choose $P_{\rm f} =0.01, P_{\rm d}=0.99$, 
$\rho_{\rm crit}$ would be $4.65$, which is close to the threshold $5$ set in~\cite{Bhagwat:2016ntk}. Here we also pick $\rho_{\rm crit}=5$.

\vspace{0.2cm}

{\noindent}{\bf Assessing observational prospects}. To investigate the
detectability of the $33$-mode after coherent stacking we employ a
Monte-Carlo (MC) sampling of possible events, repeating each sampling
$100$ times to accumulate statistics. Given the predictions
derived from the recent GW
detections~\cite{TheLIGOScientific:2016pea}, we assume a uniform
merger rate of quasi-circular inspirals of $40 \; {\rm Gpc}^{-3} {\rm
  yr}^{-1}$ in co-moving volume. For simplicity we assume the BHs are
non-spinning (see the Supplemental Material for the effect of BH spins
on the relative phase difference between the 22- and 33-mode) with
masses uniformly distributed $\in [10-50]M_{\odot}$, and employ the
empirical fitting formula of~\cite{Husa:2015iqa} to connect the
initial BH masses to the final mass and spin of the remnant.  We
compute the total SNR for each individual event using the sky-averaged
IMRPhenomB waveform model~\cite{Ajith:2009bn}, choose the amplitude of
the primary 22-mode to match the ringdown SNR in Eq.~(1)
of~\cite{Berti:2016lat}, and set the amplitude of the $33$-mode
following the fitting formula for $A_{33}/A_{22}$
in~\cite{London:2014cma}. We adopt the zero-detuned, high-power noise
spectral density of aLIGO at design sensitivity~\cite{Ajith:2011ec}
for $S_{n_j} (f)$. For each MC sampling we randomly distribute merger
events within redshift $z=1$ (with $H_0 =70\, {\rm km}\, {\rm s}^{-1}
{\rm Mpc}^{-1}, \Omega_m=0.3$) over a one year observation period, and
as discussed earlier only select those with $\rho_{22}>8$. Each MC
sampling contains about $1000-2000$ events, giving rise to $40-65$
events with $\rho_{22}>8$, which is roughly two times higher than
samples taken using population synthesis
models~\cite{Berti:2016lat}. In computing the stacked signal
SNR of Eq.~\eqref{eq:rho33}, it suffices to use a small number (15) of
loudest events in each sample~\footnote{The choice of 15 loudest events
  was to lower the computational cost of the optimization, and future
  work will investigate the optimal choice of $N$ to balance
  minimizing computational cost vs maximizing SNR}, and we determine the
weight constants $c_j$ in the sum to maximize the SNR using the
downhill simplex optimization method~\cite{Nelder1965, Press:2007nr}.

\begin{figure}[tb]
\includegraphics[width=7.4cm]{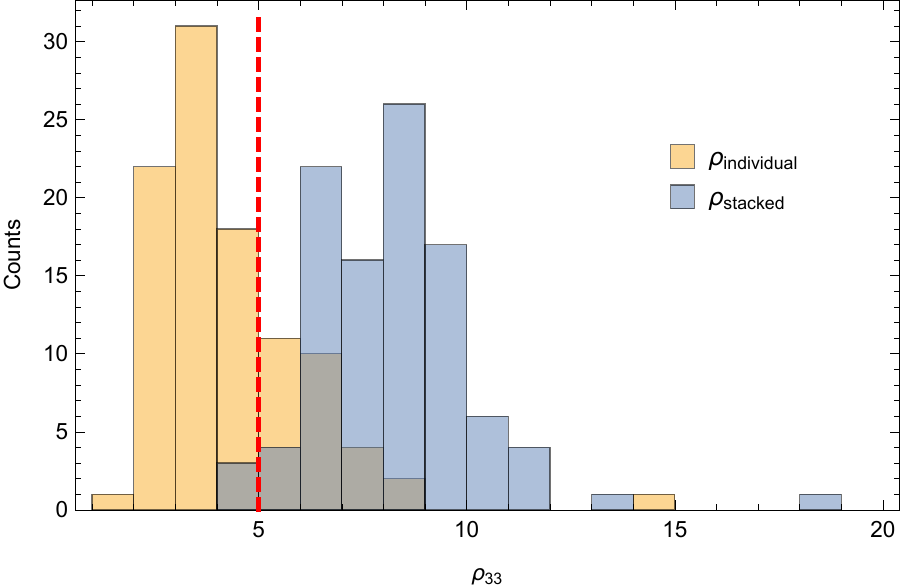}
\caption{A histogram of the SNR of the 33-mode, $\rho_{33}$, from $100$ randomly sampled sets of detections, assuming a one year data acquisition time for aLIGO and uniform co-moving merger rate of $40 \; {\rm Gpc}^{-3} {\rm yr}^{-1}$~\cite{TheLIGOScientific:2016pea}. We present the statistics of the largest $\rho_{33}$ event from each set (orange bins), and those with the stacked SNR using only the 15 largest SNR events from each set (blue bins). The 33-mode is detected if $\rho_{33}$ is above the detection threshold of $\rho_{33}=5$ (red dashed line). Refer to the main text for more details. }
\label{fig:plot1} 
\end{figure}

The resulting distribution (Fig.~\ref{fig:plot1}) indicates that there is roughly a $28\%$ chance for aLIGO to resolve at least one $33$-mode from a single event in one year of data at design sensitivity. After stacking, the probability of a collective 33-mode detection increases to $97\%$. These probabilities of course depend on the actual merger rate, as well as additional factors we have not taken into account here, including initial BH spins and precession. For example, if we take the more pessimistic event rate estimate of $13 \; {\rm Gpc}^{-3} {\rm yr}^{-1}$~\cite{TheLIGOScientific:2016pea}, the probability for detection with a single event drops to $\sim 12\%$, while the collective mode detection probability drops to $50\%$ (still using 15 events).

In theory, all else being equal, coherent stacking should provide a $\sqrt{N}$ scaling of the SNR. Here $N = 15$, so the ideal scenario would see a factor $\sim 3.8$ improvement in the collective $\rho_{33}$ relative to a single event. In our MC realizations we achieved improvement factors of between $1.3$ and $3.1$ relative to the loudest event over the set of 100 realizations (see the Supplemental Material for some additional comments and figures about the distribution). The primary reason for this is simply the non-uniform nature of the sampling, where it is typically the small handful of loudest events that contribute most to the collective SNR. The parameter uncertainty noise has smaller impact, in particular because the fainter events that have larger uncertainties are weighted less in the sum. 

\vspace{0.2cm}

{\noindent}{\bf Discussion}. We have presented a coherent mode stacking method that uses multiple high quality BBH coalescence detections to obtain better statistics for BH spectroscopy. Crucial to the method's success is the appropriate alignment of the phase and frequency from different signals. For the class of BBH merger events we have targeted here, this is achievable for two primary reasons: (1) the no-hair properties of isolated BHs in GR imply that a binary system is likewise described by a small set of parameters, (2) the expected events that aLIGO will detect where the primary ringdown mode is visible will also have an inspiral detectable with high SNR, and this can be used to estimate the parameters in (1) with enough accuracy to predict the initial phases and amplitudes of sub-dominant ringdown modes. In this first, proof-of-principle study, we have demonstrated that detection of a collective secondary BH ringdown mode through stacking is likely with the \emph{current} ``advanced'' generation of ground-based GW detectors, even if the corresponding modes are not loud enough to be detected in any single-event analysis. 

There are many avenues for future work and extensions of this method,
including using merger rates predicted by population synthesis models
as done in~\cite{Berti:2016lat}, considering other ringdown modes
(such as the 44- and 21-modes~\cite{Berti:2016lat,Bhagwat:2016ntk}, or
even the fundamental 22-mode in a population of low SNR events where
it is not individually detectable), adding spin to the progenitor BHs
and also targeting secondary inspiral modes. Furthermore, this method
could be adapted to constrain or search for other small-amplitude
features that might be shared by a population of events, e.g.~common
parameterized post-Einsteinian-like~\cite{ppe} corrections to the
inspiral phase of the mergers, or common
equation-of-state-discriminating frequencies excited in hypermassive
remnants of binary neutron star mergers
\cite{Stergioulas:2011gd,Takami:2014zpa,Takami:2014tva,Bauswein:2015vxa,Bauswein:2015yca,Paschalidis:2015mla,East:2015vix,East:2016zvv,Lehner:2016lxy,Radice:2016gym,Lehner:2016wjg}. In
this latter example, one issue in adapting the coherent stacking
method would be achieving phase alignment, due to the challenge in
accurately calculating the details of the matter dynamics
post-merger. If the phases cannot be aligned, incoherent power
stacking could still in theory achieve a $N^{1/4}$ SNR scaling (see
Supplemental Material for more details).

{\it Acknowledgements-} HY thanks Haixing Miao for sharing the code
for downhill simplex optimization. The authors thank Emanueli Berti, Swetha Bhagwat, Vitor Cardoso,
Neil Cornish, Kendrick Smith, Chris Van Den Broeck and John Veitch 
for valuable discussions and comments.
K.Y.~acknowledges support from JSPS Postdoctoral Fellowships for
Research Abroad.  F.P. and V.P.~acknowledge support from NSF grant
PHY-1607449 and the Simons Foundation. V.P. also acknowledges support
from NASA grant NNX16AR67G (Fermi). N.Y.~acknowledges support from NSF
CAREER Grant PHY-1250636. Computational resources were provided by
XSEDE/TACC under grant TG-PHY100053.  This research was supported in
part by NSERC, and in part by the Perimeter Institute for Theoretical
Physics.  Research at Perimeter Institute is supported by the
Government of Canada through the Department of Innovation, Science and
Economic Development Canada, and by the Province of Ontario through
the Ministry of Research and Innovation.

\newpage

\section{Supplemental Material}

\subsection{Details in deriving the hypothesis test}

The Generalized Likelihood Ratio Test (GLRT) was first presented in~\cite{2005ITSP...53.2579S} and applied to the ringdown analysis in~\cite{Berti:2007zu} for single detection cases in the time domain, assuming white noise. Here we apply the same technique for the stacked signals we consider in this paper and we work in the frequency domain to account
for the fact that detector noise is not white. We also include the effect of the parameter estimation noise due to the dominant mode subtraction in the analysis. 

Let us start with the probability function (Eq.~\eqref{eq:prop-PA} in the main text) 
\begin{eqnarray}
\label{eq:Likelihood}
{\rm P}_{A} &\propto& {\rm exp} \left [ -\int^\infty_0 df \frac{2 |\tilde y-A \tilde \hS_{33}|^2}{S_n}\right ]\,,\nonumber \\
 &\propto& \prod_{f>0} {\rm exp} \left [ -\frac{2 |\tilde y-A \tilde \hS_{33}|^2}{S_n}\right ]\,,
 \end{eqnarray}
 where the second line gives the discrete expression for ${\rm P}_{A}$
 and the product $\prod$ is over different frequency bin
 contributions. By extremizing the likelihood, the maximum likelihood
 estimator for the amplitude is
 \begin{align}
 \hat{A} = & \frac{\langle \hS_{33} | y \rangle }{\langle \hS_{33} | \hS_{33} \rangle} 
 = \frac{1}{2} \frac{\int^\infty_0 df \frac{ \tilde h^*_{33}(f) \tilde y(f) + \tilde h_{33}(f) \tilde y^*(f) }{S_n(f)}}{\int_0^\infty df \frac{|\tilde \hS_{33}|^2}{S_n}} \,,
 \end{align}
with
 \begin{align}
 \langle \chi | \xi\rangle \equiv 2 \int^\infty_0 \frac{\tilde \chi^* \tilde \xi +\tilde \chi \tilde \xi^*}{S_n} df.
 \end{align}
 
 In order to perform the GLRT test, we compute the following quantity 
 \begin{align}
 \label{eq:T_y}
 T(y) = \ln \frac{{\rm max}_{\mathcal{H}_1} P_{A}}{{\rm max}_{\mathcal{H}_2} P_{A=0}}= \frac{\hat{A}^2}{2} \langle \hS_{33} | \hS_{33} \rangle\,,
 \end{align}
where in our specific situation, ${\rm max}_{\mathcal{H}_1}P_{A} = \max_A P_A$ and ${\rm max}_{\mathcal{H}_2}P_{A=0} = \max_A P_{A=0} = P_{A=0}$.
Notice that since $P_{A=0}$ for hypothesis 2 does not depend on $A$, its maximization over $A$ simply gives $P_{A=0}$ itself.
 Assuming that the noise is Gaussian, $\sqrt{2 T(y)}$ also follows a Gaussian distribution and
one can propose that hypothesis $1$ is preferred if
 \begin{align}
 \label{eq:T_y-threshold}
 \sqrt{2 T(y)} =\frac{\langle \hS_{33} | y\rangle }{ \sqrt{\langle \hS_{33} | \hS_{33} \rangle}} = \frac{\langle \hS_{33} | y\rangle }{ ||\hS_{33}||}  >\Gamma_1\,.
 \end{align}
Here, $\Gamma_1$ is defined as $\Gamma_{\sigma^2}$ with the variance $\sigma^2=1$, where $\Gamma_{\sigma^2}$ is given by the false-alarm rate $P_f$: $\Gamma_{\sigma^2} = Q_{\sigma^2}^{-1}(P_f)$ with $Q_{\sigma^2}(x)$ representing the right-tail probability function for a Gaussian distribution with zero mean and variance $\sigma^2$:
 \begin{align}
 Q_{\sigma^2}(x) \equiv \frac{1}{\sqrt{2\pi} \sigma} \int^\infty_x e^{-\frac{z^2}{2\sigma^2}} dz\,.
 \end{align}   
 The noise component of Eq.~\eqref{eq:T_y-threshold} is a normalized Gaussian distribution with zero mean and unit variance, with the latter explicitly given by
 \begin{widetext}
\allowdisplaybreaks
\begin{align}
{\rm Var}\left [ \frac{\langle \hS_{33} | \nS\rangle }{ || \hS_{33}||}\right ] &= \left \langle \left( \frac{\langle \hS_{33} | \nS\rangle }{ || \hS_{33}||} \right)^2 \right \rangle - \left \langle \frac{\langle \hS_{33} | \nS\rangle }{ || \hS_{33}||}  \right \rangle^2 \nonumber \\
&=\left \langle \frac{4}{|| \hS_{33}||^2} \int^\infty_0\int^\infty_0 df df' ( \tilde \hS^*_{33}(f) \tilde n(f)+h.c.) ( \tilde \hS^*_{33}(f') \tilde n(f')+h.c.) \frac{1}{S_n(f) S_n(f')} \right \rangle \nonumber \\
&= \frac{4}{|| \hS_{33}||^2} \int^\infty_0 df  \frac{|\tilde \hS_{33}(f)|^2}{S_n(f)}=1\,.
\end{align}
\end{widetext} 
Here we used
\allowdisplaybreaks
\begin{align}
\langle n (f) \rangle&=0, \\
 \langle n (f) n(f') \rangle&=0, \\
 \langle n(f)  n^*(f') \rangle &=\frac{1}{2} S_n(f) \delta(f-f')\,,
\end{align}
for one-sided spectrum $S_n$ with $\langle X \rangle$ representing the
expectation value of $X$, and the averaging operation $\langle
\rangle$ is defined over an ensemble of noise realizations.

At this point, we notice that we only know the maximum likelihood estimator $y'$ instead of $y$ (recall $y'= y + \delta y)$ . In particular ~\footnote{One ends up with the same expression even if one introduces the probability distribution of $n_h$ in Eq.~\eqref{eq:Likelihood} (which cancels in $T(y')$) and use $\langle n_{h_{22}} \rangle$=0.},
\begin{align}
\label{eq:Ty'}
\sqrt{2 T( y')} &\approx \frac{\langle {\hS}'_{33} | A {\hS}_{33}+\nS-n_{h_{22}} + A n_{h_{33}}\rangle }{ || {\hS}'_{33}||} \nonumber \\
&= A ||{\hS}_{33} + n_{h_{33}} || +  \frac{\langle {\hS}_{33}+n_{h_{33}} | \nS-n_{h_{22}} \rangle }{ || {\hS}_{33}+ n_{h_{33}}||}\,, 
\end{align}  
where the parameter uncertainty noise $n_{h_{\ell m}}$ is defined above Eq.~\eqref{eq:paran}.
Let us further assume that noise is small and keep up to its second order. Neglecting cross terms such as $n_{h_{33}} \nS$ and $n_{h_{33}} n_{h_{22}}$, where the former term has zero mean and the latter term is small due to the separation of resonance for $22$ and $33$ mode, the above equation becomes
\begin{align}
\sqrt{2 T( y')} & = A ||{\hS}_{33} || + A \frac{\langle {\hS}_{33} | n_{h_{33}} \rangle}{||{\hS}_{33} ||} + \frac{A}{2} \frac{\langle n_{h_{33}} | n_{h_{33}} \rangle}{||{\hS}_{33} ||} \nonumber \\
& - \frac{A}{2} \frac{\langle {\hS}_{33} | n_{h_{33}} \rangle^2}{||{\hS}_{33} ||^3}  +  \frac{\langle {\hS}_{33} | \nS-n_{h_{22}} \rangle }{ || {\hS}_{33}||} + \mathcal{O}(n^3)\,.
\end{align}  

Let us now derive a criterion for hypothesis 1 to pass the GLRT test
including the parameter estimation noise. 
In the following, we use a bar to denote quantities for hypothesis 2 (not to be confused with the averaging operator $\langle \rangle$)
 while unbarred quantities refer to those for hypothesis 1.
To $\mathcal{O}(\delta^2)$, the distribution of $\sqrt{2 T(y')}$ for hypothesis 2 ($A=0$) has mean 
  \begin{align}
\bar \mu = - \frac{\langle {\hS}_{33} | \langle n_{h_{22}} \rangle \rangle}{ || {\hS}_{33}|| }\,,
 \end{align}
 and variance
  \begin{align}\label{eq:barGamma}
& {\rm Var}\left [ \frac{\langle \hS_{33} | \nS-n_{h_{22}}\rangle }{ || \hS_{33}||}\right ] \nonumber \\
&=1+{\rm Var}\left [ \frac{\langle \hS_{33} | n_{h_{22}}\rangle }{ || \hS_{33}||}\right ] \equiv 1+\bar \sigma^2_{\rm p}\,,
 \end{align}
 where we neglect the correlation between $\nS$ and $n_{h_{22}}$.
Let us next shift the distribution by $- \bar \mu$ such that the shifted distribution has zero mean and
denote the right-tail probability of the shifted distribution above $x$ as $\bar Q(x)$. 

Although the distribution is not a Gaussian due to $\mathcal{O}(\delta^2)$ terms in $\sqrt{2T(y')}$, we next show that 
only the Gaussian part of $\bar Q(x)$ contributes to $\bar \Gamma \equiv \bar Q^{-1}(P_f)$. 
We start by noting that $\bar Q$ is 
the right-tail probability of $\langle \hS_{33} | \nS \rangle/|| {\hS}_{33}|| - \langle \hS_{33} | n_{h_{22}} \rangle/|| {\hS}_{33}|| + \bar \mu$. The first and second terms are Gaussian noise to $\mathcal{O}(\delta)$ so that the sum is also Gaussian to that order, with variance being $1+\bar{\sigma}^2_{\rm p}$ and $\bar{\Gamma}=\Gamma_1 \sqrt{1+\bar{\sigma}^2_{\rm p}}$. Therefore the presence of $\mathcal{O}(\delta)$ noise component shifts $\Gamma_1$ by $\mathcal{O}(\delta^2)$ order. Because the non-Gaussian noise component in $\sqrt{2 T(y')}$ enters at $\mathcal{O}(\delta^2)$, its effect on $\bar{\Gamma}$ is at least on $\mathcal{O}(\delta^3)$, which is beyond the order of perturbation we are considering.
We conclude that we only need to consider the Gaussian contribution in $\bar Q$ to derive $\bar \Gamma$ valid to $\mathcal{O}(\delta^2)$.
 
 Thus, to the perturbation order we are working, it suffices to assume that
 the shifted distribution is a Gaussian given by $\bar Q = Q_{1+\bar \sigma^2_{\rm p}}$.
Having such $\bar Q$ at hand, 
the criterion for hypothesis 1 to be preferred over hypothesis
2 for a given $y'$ and $P_f$ given in Eq.~\eqref{eq:T_y-threshold} is modified to
\begin{align}
\label{eq:T_y-threshold-prime}
\sqrt{2 T(y')} >  \bar Q^{-1}(P_f) = \Gamma_{1+ \bar \sigma_{\rm p}^2}\,,
\end{align}
where $\Gamma_{1+ \bar \sigma_{\rm p}^2}$ is equivalent to $\bar \Gamma$ in Eq.~\eqref{eq:barGamma}.

On the other hand, the distribution of $\sqrt{2 T(y')}$ for hypothesis 1 with $A=1$ to $\mathcal{O}(\delta^2)$ has mean
\begin{align}
\mu &= || \hS_{33}||+ \frac{\langle {\hS}_{33} | \langle n_{h_{33}} \rangle \rangle}{||{\hS}_{33} ||} + \frac{1}{2} \left \langle \frac{\langle n_{h_{33}} | n_{h_{33}} \rangle}{||{\hS}_{33} ||} \right \rangle \nonumber \\
& - \frac{1}{2} \left \langle \frac{\langle {\hS}_{33} | n_{h_{33}} \rangle^2}{||{\hS}_{33} ||^3} \right \rangle \nonumber \\
&= ||{\bf H}_{33}|| + \bar \mu\,,
\end{align}
and variance
\begin{align}
& {\rm Var}\left [ \frac{\langle \hS_{33} | \nS-n_{h_{22}} + n_{h_{33}} \rangle }{ || \hS_{33}||}\right ] \nonumber \\
&=1+{\rm Var}\left [ \frac{\langle \hS_{33} | n_{h_{22}} - n_{h_{33}}\rangle }{ || \hS_{33}||}\right ] \equiv 1+ \sigma^2_{\rm p}\,,
\end{align}
where ${\bf H}_{33}$ corresponds to the reduced 33 mode signal due to parameter uncertainties and is given by 
\begin{align}
{\bf H}_{33} &= \langle \hS_{33} \rangle + \frac{1}{2} \left( \left \langle \frac{\langle n_{h_{33}} | n_{h_{33}} \rangle}{\langle {\hS}_{33} | {\hS}_{33} \rangle} \right \rangle  -  \left \langle \frac{\langle {\hS}_{33} | n_{h_{33}} \rangle^2}{\langle {\hS}_{33} | {\hS}_{33} \rangle^2} \right \rangle \right) \hS_{33} \nonumber \\
&= \left[1 + \frac{1}{2} \left( \left \langle \frac{\langle n_{h_{33}} | n_{h_{33}} \rangle}{\langle {\hS}_{33} | {\hS}_{33} \rangle} \right \rangle  -  \left \langle \frac{\langle {\hS}_{33} | n_{h_{33}} \rangle^2}{\langle {\hS}_{33} | {\hS}_{33} \rangle^2} \right \rangle \right) \right] \langle \hS_{33} \rangle \nonumber \\
& + \mathcal{O}(\delta^3)\,,
\end{align}
 with
 \begin{align}\label{eq:h33b}
\langle \hS_{33} \rangle &= e^{i (\omega_{33} t -  \phi_{33})} \sum_j  \AS_{33,j} e^{-\gammaS_{33,j} t- (\sigma_{\Lambda_{33,j}} t)^2/2-(\sigma_{\phiS_{33,j}})^2/2} \nonumber \\
 & + \mathcal{O}(\delta^3)\,.
  \end{align} 
Here we used $\langle e^x
\rangle =e^{\sigma_x^2/2}$ with the variance $\sigma_x^2 \equiv
\langle x^2 \rangle$ for any {\it complex} Gaussian random variable $x$.
Following the case for hypothesis 2, we shift the distribution by $-\mu$ such that its mean becomes zero. Then, the right-tail probability of the shifted distribution is simply given by $Q_{1+\sigma_p^2}$. Notice that the non-Gaussian contribution can be neglected as we discussed in the hypothesis 2 case. 

To claim a detection of the 33 mode, we
require that Eq.~\eqref{eq:T_y-threshold-prime} be satisfied
with the detection rate $P_d$. The criterion is given by
\begin{align}
P_d \leq  Q_{1+\sigma_p^2} \left( \Gamma_{1+ \bar \sigma_{\rm p}^2}  - \mu + \bar \mu \right)\,.
\end{align}
Using further the relation $Q_{\sigma^2}(x) = Q_1(x/\sigma)$, the above equation reduces to
\begin{align}
\frac{||{\bf H}_{33}||}{\sqrt{1+\sigma_p^2}} \geq \frac{\Gamma_{1+ \bar \sigma_{\rm p}^2}}{\sqrt{1+\sigma_p^2}} - Q_{1}^{-1}(P_d)\,.
\end{align}
The left and right hand side of this inequality correspond to the SNR of the 33 mode including parameter uncertainties and the critical SNR for detection respectively. To simplify the latter further, we choose to be more conservative and replace $\Gamma_{1+ \bar \sigma_{\rm p}^2}$ with $\Gamma_{1+ \sigma_{\rm p}^2} (\geq \Gamma_{1+ \bar \sigma_{\rm p}^2})$:
\begin{align}
\label{eq:rho_crit}
\rho_{33} \equiv \frac{|| {\bf H}_{33}||}{\sqrt{1+\sigma^2_{\rm p}} }\ge Q_1^{-1}(P_f) -Q_1^{-1}(P_d) \equiv \rho_{\rm crit}\,,
\end{align}
where we used $\Gamma_{\sigma^2} = \sigma \Gamma_1$.
   
\subsection{Estimating uncertainties in target mode phase}
Of all the parameters considered in this work, the accuracy in estimating the
constant phase offsets $\phi_{33,j}$ is the most important in improving
the collective SNR. Here, we discuss in more detail the two dominant
sources of error in this quantity. 

The first comes from uncertainties in the intrinsic parameters
estimated from each event, including the masses and spins of the
individual BHs prior to merger. We estimate this effect in the
following way.  First, we employ full inspiral-merger-ringdown
waveforms obtained with a numerical relativity surrogate
model~\cite{Blackman:2015pia,Blackman:2017dfb} with which we produce different
waveforms to measure the individual {\em total phases}
$\bm{\Phi\mkern-11mu\Phi}_{33}\equiv \omega_{33} t + \phi_{33}$ and
$\bm{\Phi\mkern-11mu\Phi}_{22}\equiv \omega_{33} t + \phi_{22}$.
Next, we time-shift the signals so that $t=0$ corresponds to the maximum
amplitude of the GW. The difference between these phases is shown at
the top plot of Fig.~\ref{fig:plot2} for representative values of the
mass ratio in binaries. Notice then that at $t=0$ one has a measure of
$\phi_{33}$ relative to $\phi_{22}$. (Also, since the instance at
which $t=0$ is chosen and the onset of the QNM is not sharply defined,
we show the phases within a time-window around the peak in GW
amplitude).  To assess how this phase difference changes for different
BH masses and spins, we vary these values within the uncertainties
reported for GW150914 and plot the difference $\bm\Delta
\bm{\Phi\mkern-11mu\Phi}_{33-22}$ in the middle and bottom panels of
Fig.~\ref{fig:plot2}~\footnote{We have also run numerical relativity
  simulations with the code of~\cite{Etienne:2007jg,Gold:2014dta} and
  confirmed that the results in Fig.~\ref{fig:plot2} are consistent
  with the simulations. }.

An additional possible source of uncertainty in $\phi_{33,j}$ is due to uncertainties in the polarization and inclination angles of the source 
(relative to the line of sight), as the dependence of $\phi_{33,j}$ on the  polarization phase can be different among different $\ell$ modes. 
However, such uncertainties are of order $\sim 1\%$. This can be seen by noticing that in a spin-weighted spherical harmonic decomposition no differences
arise~\cite{Klein:2009gza} and the transformation to the required spin-weighted spheroidal harmonic introduce such small effect~\cite{Berti:2005gp,Berti:2007zu,PhysRevD.90.064012}. 
Thus, this source of uncertainty is negligible in our analysis.

\begin{figure}[tb]
\includegraphics[width=8.5cm]{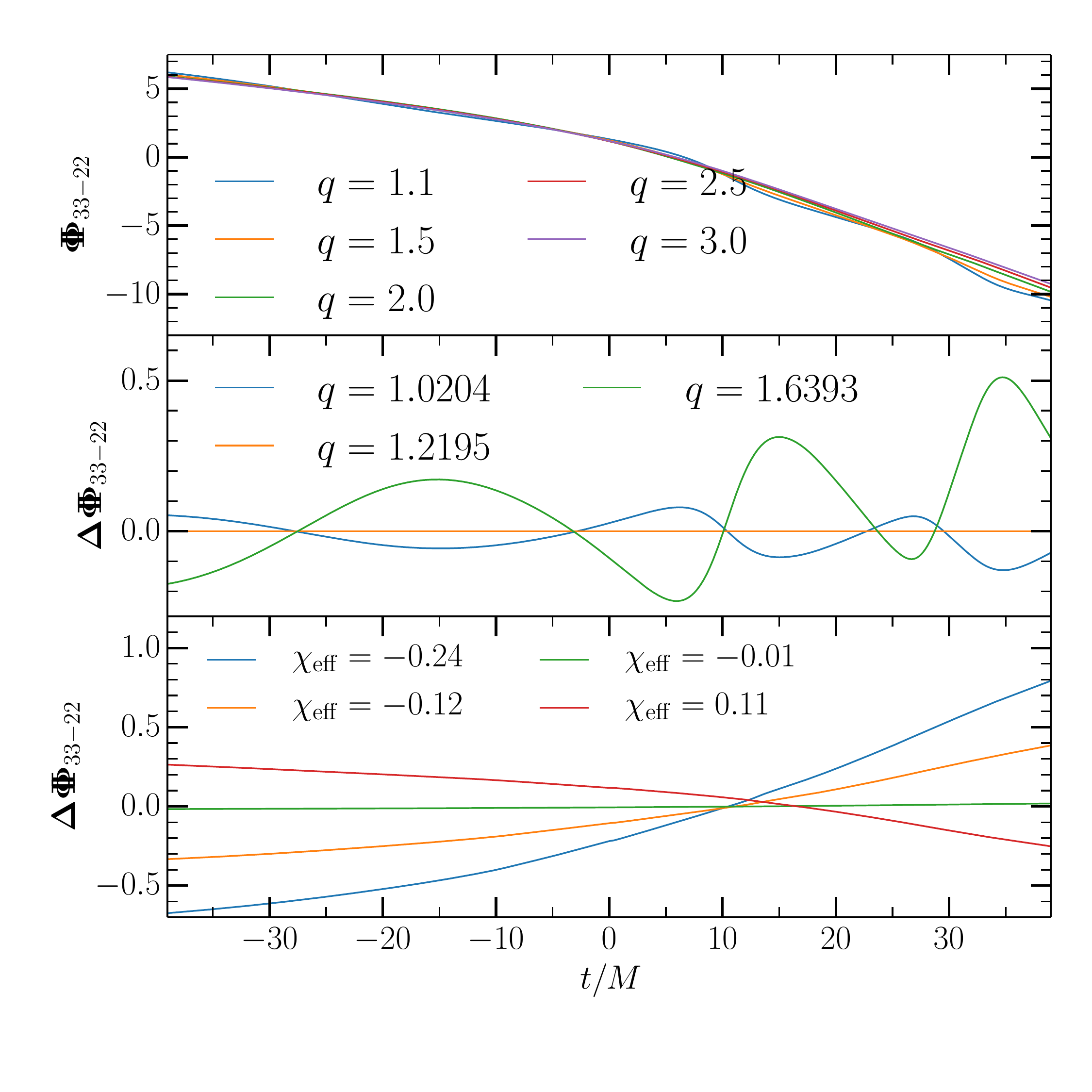}
\caption{Top panel: difference between the total phase of the 33 mode
  and that of the 22 mode, i.e.,
  $\bm{\Phi\mkern-11mu\Phi}_{33-22}=\bm{\Phi\mkern-11mu\Phi}_{33}-\bm{\Phi\mkern-11mu\Phi}_{22}$
  for different mass ratios.  Middle/Bottom panel: variation of
  $\bm{\Phi\mkern-11mu\Phi}_{33-22}$ using the expected mean parameters
  of ${\rm GW150914}$ and expected mass ratio/effective spin
  uncertainties of ${\rm GW150914}$~\cite{PhysRevLett.116.241102} (to
  $90\%$ credible levels) . Here the spins of both black holes are
  assumed to be aligned and equal (other spin combinations within the
  confidence interval of $\chi_{\rm eff}$ give similar variations). }
\label{fig:plot2}
\end{figure}

Based on the above considerations, we estimate $\sigma_{\phiS_{33}} =
0.3 \times (20/ \rho)$ rads, where the value of 0.3 rads for $\rho =
20$ is extracted from the middle and bottom panels of
Fig.~\ref{fig:plot2} with $t \in (0,10)M$, within which we expect the
onset of the ringdown phase. While this estimate is obtained from GW$150914$, we anticipate that generally BH binaries could have very different spin configurations. Understanding the spin dependence of phase errors is necessary for more systematic future studies. The $1/\rho$ scaling can be obtained
through a straightforward Fisher analysis and error propagation as
follows: Using an IMR waveform, we estimate the covariance matrix
$\Sigma_{ab}^\mathrm{(insp)}$ of the inspiral parameters (individual
masses and spins) $\theta^a_\mathrm{(insp)}$ as the inverse of the
Fisher matrix. Since $\theta^a_\mathrm{(insp)}$ are related to the
ringdown parameters $\theta^a_\mathrm{(rd)} = (A_{\ell m},
\Omega_{\ell m}, \Gamma_{\ell m}, \Phi_{\ell m})$, we approximately
obtain the covariance matrix of the latter as
\begin{align}
\Sigma_{ab}^\mathrm{(rd)} = \sum_{p,q} \frac{\partial \theta^a_\mathrm{(rd)}}{\partial \theta^p_\mathrm{(insp)}} \frac{\partial \theta^b_\mathrm{(rd)}}{\partial \theta^q_\mathrm{(insp)}} \Sigma_{pq}^\mathrm{(insp)}\,.
\end{align}
Since $\Sigma_{ab}^\mathrm{(insp)}$ is proportional to $1/\rho^2$, the
uncertainty in $\theta^a_\mathrm{(rd)}$ (equivalent to
$\sqrt{\Sigma_{aa}^\mathrm{(rd)}}$) scales as $1/\rho$. One can
also use the previous formula to estimate the amount of correlation
among the ringdown parameters.

\subsection{Monte-Carlo sampling, comparison to earlier single-rate estimates, and SNR boost through stacking}

Here we provide some additional comments and details regarding our Monte-Carlo
sampling of simulated events, illustrated in Fig.~\ref{fig:plot1} in the main text, and
Fig.~\ref{fig:plot3} below. 

First, our estimate of a 0.3/yr detection rate for the 33-mode without the
coherent mode stacking implied in Fig.~\ref{fig:plot1} is larger than the $\sim 0.03$/yr rate
predicted in~\cite{Berti:2016lat}. One of the reasons for this difference
arises from the value of $A_{33}/A_{22}$ we have used. Here we employ
the fitting formula derived in~\cite{London:2014cma}, which
typically gives a ratio 1.6 times larger than that used in the earlier
study. The difference between the ratios from these fitting formulas is mostly related to the choice of ``starting time" of QNMs. Had we instead used the ratio as in~\cite{Berti:2007zu,Berti:2016lat},
it would have effectively raised $\rho_\mathrm{crit}$ to $4.65 \times 1.6
\sim 7.5$, dropping the expected event rate of the 33-mode to $\sim
0.06$/yr (see Fig.~\ref{fig:plot1}). The second reason for our higher
rate comes from the larger merger rate of  $40 {\rm Gpc}^{-3} {\rm yr}^{-1}$~\cite{TheLIGOScientific:2016pea}
that we use. These two factors together make our single-event
rate estimate consistent with~\cite{Berti:2016lat}.

As mentioned in the main text, the reason we do not get prefect 
$\sqrt{N}$ scaling when stacking is due to the non-uniform distribution
of SNRs. In a typical sample, the individual SNRs have a pyramid-like 
distribution (as indicated in Fig.~\ref{fig:plot3}),
and the top few loudest events matter the most enhancing the collective
vs. single-loudest event SNR. This is also why increasing the number
of events used beyond the $N=15$ chosen here will not significantly
increase the stacked SNR, and we could probably have used even fewer
than $15$ without much degradation of the SNR. 
The value
of $15$ was chosen simply to reduce the computational cost of the
simulations, and we leave it to future work to find an adequate
$N$ giving most of the SNR with least computational cost.

For illustrative purposes, in Fig.~\ref{fig:plot4} we show
that if we did have a set of identical sources we would
obtain $\sqrt{N}$ scaling in the stacking process.
There, we took $15$ events that are identical to
GW$150914$, all with the same noise spectrum, and then stacked them 
coherently as discussed in the main text.  In the figure we show the original signal,
detector noise (assuming aLIGO noise) versus the stacked signal and
stacked detector noise.

A relatively minor factor in reducing the efficacy of stacking
can be attributed to the frequency
rescaling of the noise spectrum $S_n$. Because the detector noise
curve is not flat in frequency, overlapping rescaled noise spectra can
add low-sensitivity regions to high-sensitivity ones, leading to worse
overall noise performance when compared to the case where no rescaling
is required. This could be mitigated to some extent by a judicial
choice of the particular target-mode frequency we choose to scale all events 
to; we leave that to future work to investigate.

A final adverse affect on the stacked SNR we note is due to parameter estimation noise;
we estimate it reduces the final SNR by $\sim 5\%$ in a typical MC simulation set. If future parameter uncertainty studies suggest larger phase errors (for example, imagine spin effects to be very different from GW$150914$), a more conservative estimate with $\sigma_{\phiS_{33}} =
0.6 \times (20/ \rho)$ rad (twice as we have assumed in the main text) reduces the final SNR by $\sim 15\%$.

\begin{figure*}[tb]
\includegraphics[width=8.4cm]{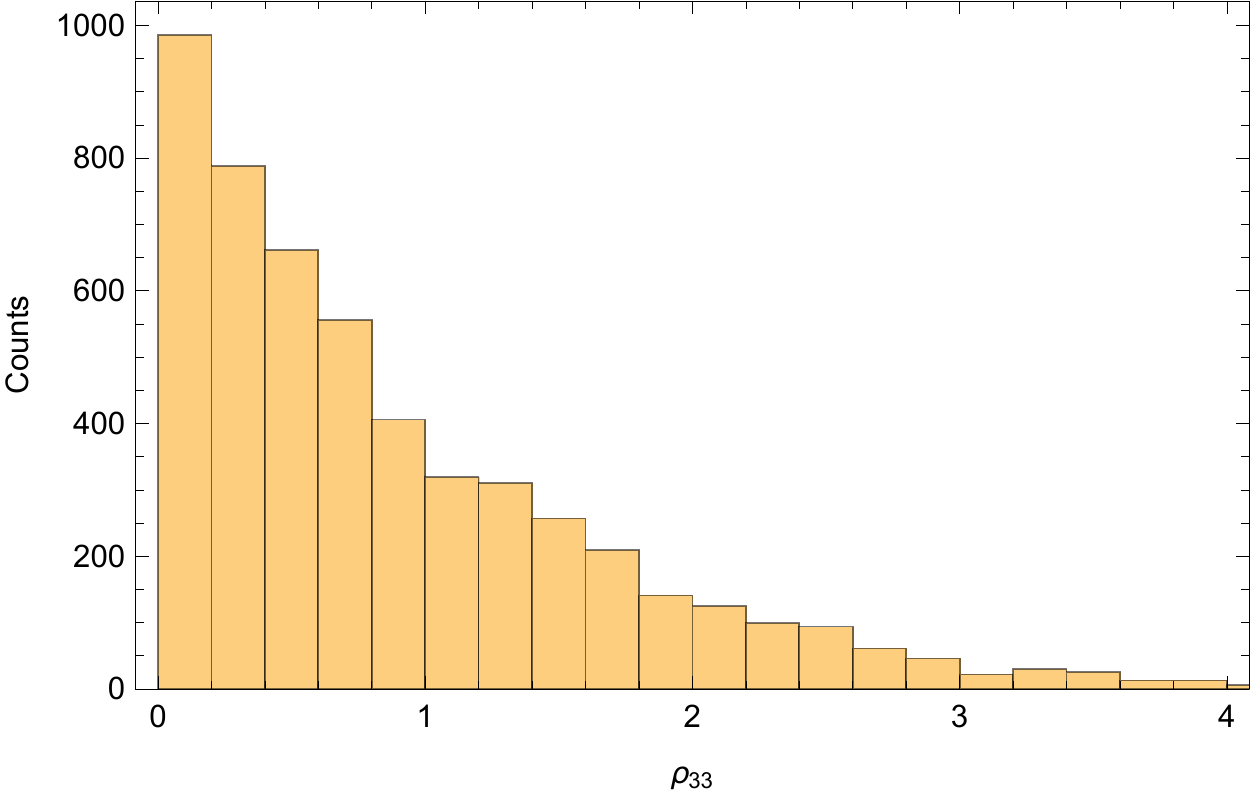}
\includegraphics[width=8.4cm]{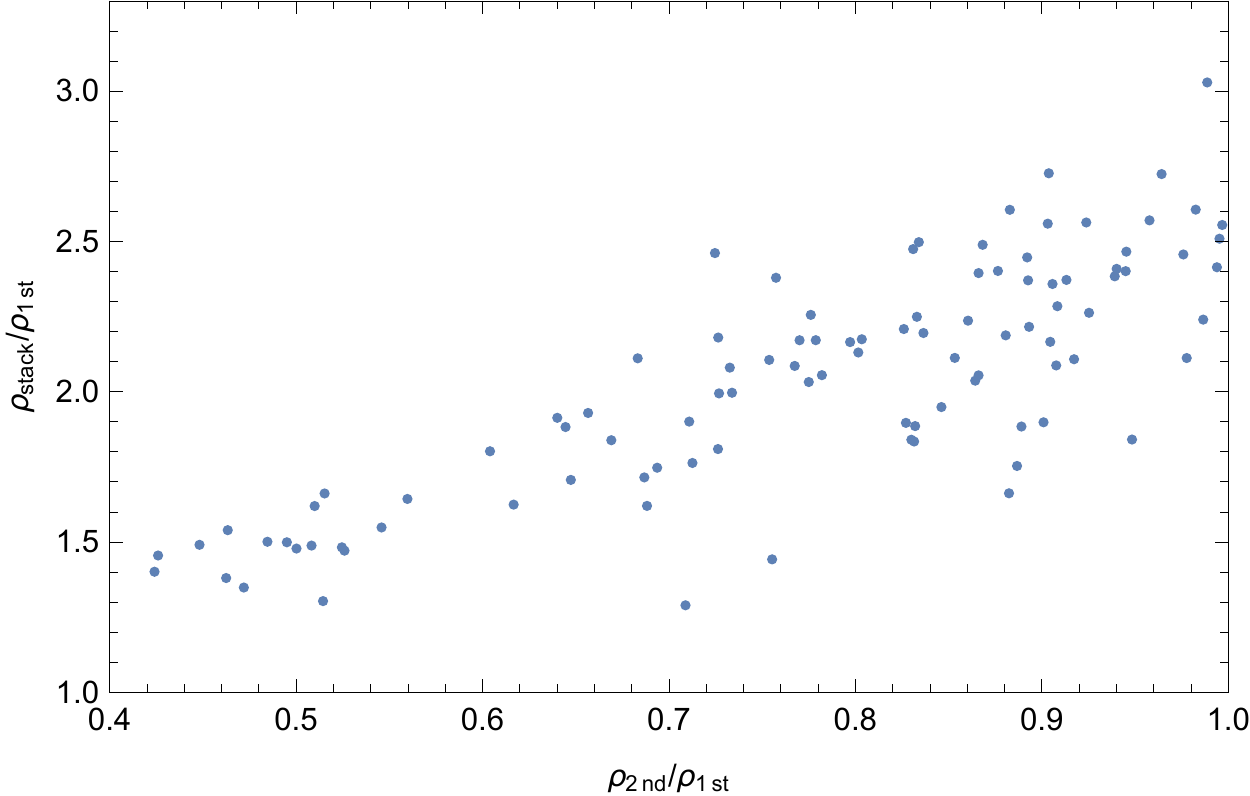}
\caption{(Left) A collection of all events in $100$ sets of data, showing the distribution of individual 33-mode SNR. The tail of the distribution does contain events with $\rho_{33}>4$ (see Fig.~\ref{fig:plot1}), though here we focus on the dominant range of the distribution. (Right) Scatter plot of the $100$ sets of data, with the horizontal axis being the ratio of $\rho_{33}$ between the second loudest and loudest event within each set, and the vertical axis being the ratio between the SNR of the stacked signal and that of the  loudest event. Observe that the coherent mode stacking works more efficiently when the SNR of the loudest event is closer to that of the second largest event.}
\label{fig:plot3}
\end{figure*}

  \begin{figure}[tb]
\includegraphics[width=8.5cm]{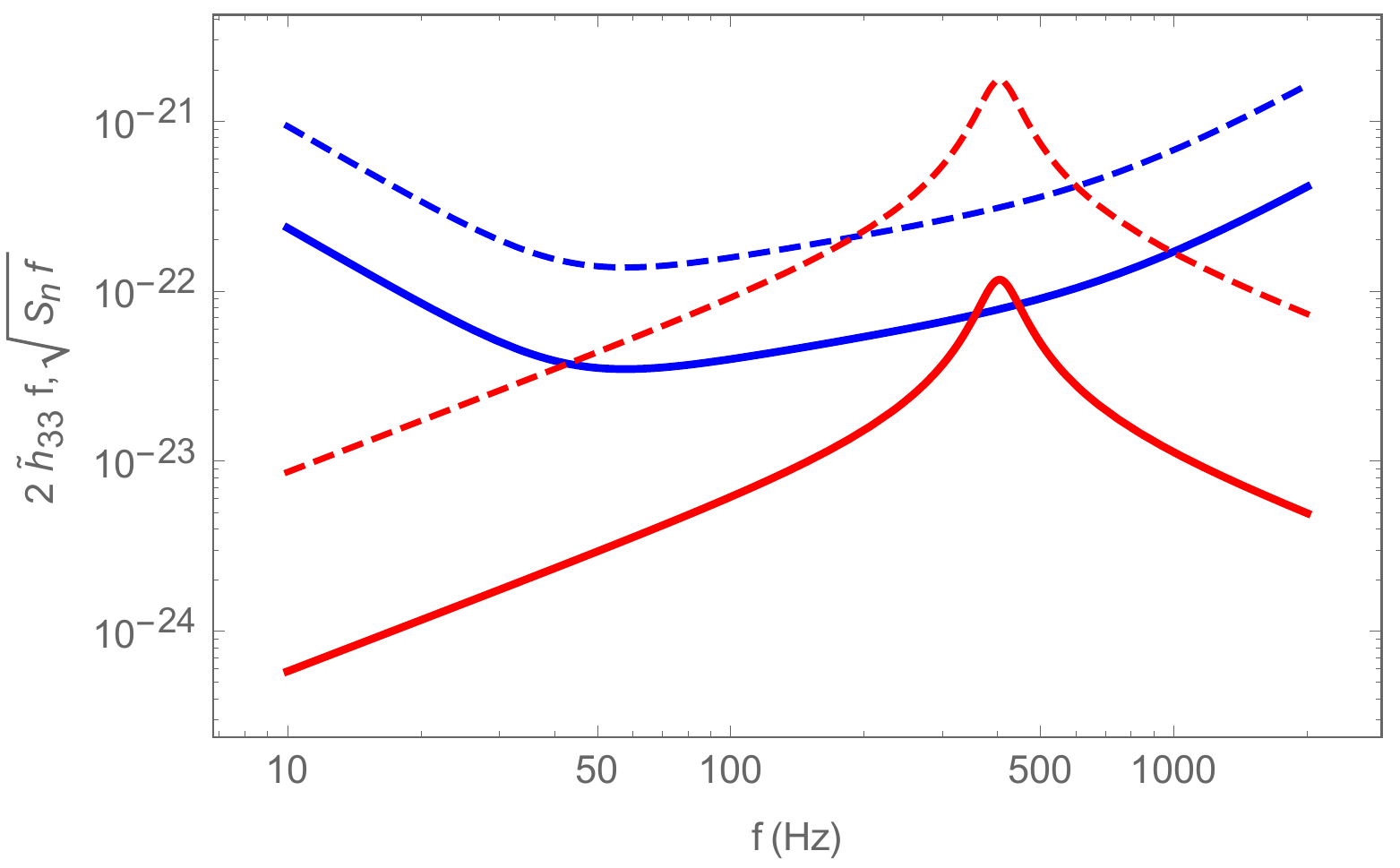}
\caption{Detector noise (thick blue), stacked noise (dashed blue)
  v.s. signal (thick red) and the stacked signal (dashed red),
  assuming $15$ ${\rm GW150914}$-like events with aLIGO
  sensitivity. For comparison purposes, we have rescaled the noise
  component and signal component so that they are both
  dimensionless. In particular, the noise is rescaled as $\sqrt{S_{n}
    f}$ and the signal is rescaled as $2 f \tilde{h}_{33}$. The stacked
  detector noise is $\sqrt{15}$ times larger than the noise of a
  single detector and the stacked signal is 15 times larger than the
  original signal.}
\label{fig:plot4}
\end{figure} 

\subsection{Power stacking}

For completeness, we note an alternative approach to stacking signals
in the hypothesis test set-up~\cite{Meidam:2014jpa}. Assuming one does
{\em not} have prior information about the phase of the $33$-mode, one
can multiply the probability function (or Bayes factors) of single
detections to obtain the total probability function
 \begin{align}\label{eq:ps}
 P_{A} \propto \prod_j \prod_{f>0} \exp \left [ -\frac{2 |\tilde{y}_j-A_j \tilde{h}_{33,j}|^2}{S_{n_j}}\right ]\,,
 \end{align}
where $j$ labels the individual detections. Each event has its own maximum likelihood estimator as given in the single detection case. The generalized likelihood ratio test suggests
 \begin{align}
 \label{eq:T_yj}
 T(y_j,j=1\ldots N)& = \ln \frac{{\rm max}_{H_1} P_{A}}{{\rm max}_{H_2} P_{A=0}}\nonumber \\
 &= \sum^N_{j=1} \frac{\hat{A}_j^2}{2} \langle h_{33,j} | h_{33,j} \rangle\,.
 \end{align} 
It is straightforward to see that the noise part of $T$ follows a $\chi^2_N$ distribution, which we label as $R$ here. We say hypothesis $1$ is preferred if 
\begin{align}
 P_d \le R \left (R^{-1}(P_f)-\sum^N_{j=1} A_j^2 \langle h_{33,j} | h_{33,j} \rangle  \right )\,,
 \end{align}
 or equivalently
 \begin{align}
 \label{eq:power-stacking-threshold}
 \sum^N_{j=1} A_j^2 \langle h_{33,j} | h_{33,j} \rangle \ge R^{-1}(P_f)-R^{-1}(P_d)\,.
  \end{align}
  
Let us assume that we are looking at events all with the same SNR. When $N$ is large, the $\chi^2_N$ distribution can be well approximated by a Gaussian distribution, so that the right hand side of the above equation scales as $\sqrt{N}$. On the other hand, the left hand side of the equation scales as $N$. As a result, the improvement due to this stacking process is equivalent to lowering $S_n$ (which comes from $\langle h_{33,j} | h_{33,j} \rangle$) by a factor $\sqrt{N}$, or the ``amplitude" of noise (characterized by $\sqrt{S_n}$) by a factor of $N^{1/4}$. Therefore this power stacking process improves the SNR with a suboptimal $\mathcal{O}(N^{1/4})$ when $N$ is large but, as described, does not require phase knowledge. Such a scaling in SNR is consistent with that in e.g.~\cite{Kalmus:2009uk}.

We now compare the previous calculations of power stacking and
coherent stacking with a Bayesian model selection study with multiple
events performed in~\cite{Meidam:2014jpa}. In this reference, the
authors construct an odds ratio of multiple events by multiplying the
Bayes factor of each event. This gives a factor of $N$ improvement on
the odds ratio compared to a single event case, just like the log of
the maximum likelihood ratio $T(y_j)$ in Eq.~\eqref{eq:T_yj} improves
by the same factor. One then needs to compare the odds ratio with a
threshold to determine which hypothesis is preferred. Since the
threshold on the right hand side of
Eq.~\eqref{eq:power-stacking-threshold} scales with $\sqrt{N}$, we
expect that the same scaling holds for the threshold of the odds
ratio. Thus, $\rho^2$ scales with $N/\sqrt{N}=\sqrt{N}$ in this case.
On the other hand, if one uses the coherent mode stacking,
$T(y)$ in Eq.~\eqref{eq:T_y} also scales with a factor of $N$ but the
threshold (corresponding to $\Gamma^2/2$ from
Eq.~\eqref{eq:T_y-threshold}) is independent of $N$. 
Thus, $\rho^2$ scales with $N$ in the coherent mode stacking case.
This is why the
coherent mode stacking should have an advantage over the power
stacking, but at the price of using full waveform
information.

At this stage, we recall that if we know the exact phase and frequency of $33$ modes in each detection {\it a priori}, or if we are performing parameter estimation for a universal parameter (let's say $A$), we can replace all $A_j$'s in Eq.~\eqref{eq:ps} by a single parameter $A$ and perform the GLRT again. In this case, a straightforward calculation shows that we gain order $\sqrt{N}$ in SNR using Bayesian approach.  Of course in reality the phase and frequency of $33$ modes are never known perfectly, but one can imagine that an improved Bayesian approach, for example using  the Bayesian model selection with a combined odds ratio
in~\cite{Meidam:2014jpa} and taking into account {\it prior} information with parameter uncertainties,  should give consistent result with the coherent mode stacking method discussed here. In other words, it is likely that the full waveform
  information can be folded into a Bayesian model selection in which
  case the improvement should be comparable to the coherent stacking
  method.
  
\bibliography{master}
\end{document}